\documentclass{PoS}

\title{Tests for the Expansion of the Universe}

\ShortTitle{Tests for Expansion}

\author{\speaker{Mart\'in L\'opez-Corredoira}%
\\
        Instituto de Astrof\'isica de Canarias, E-38205 
           La Laguna (Tenerife), Spain \\
        Universidad de La Laguna, Dept. Astrof\'\i sica, E-38206
	   La Laguna (Tenerife), Spain \\
        E-mail: \email{martinlc@iac.es}}

\abstract{
Almost all  cosmologists accept nowadays that the redshift of the galaxies is due to the expansion 
of the Universe (cosmological redshift), plus some Doppler effect of  peculiar motions, but can we be
 sure of this fact by means of some other independent cosmological test? Here I will review some 
 recent tests: CMBR temperature versus redshift, time dilation, the Hubble diagram, the Tolman or surface brightness test,
the angular size test, the UV surface brightness limit and the Alcock--Paczy\'nski test. Some tests favour
 expansion and others favour a static Universe.
 Almost all the cosmological
tests are susceptible to the evolution of galaxies and/or other effects.
Tolman  or angular size tests need to assume very strong evolution of  galaxy sizes to fit the data with 
the standard cosmology, whereas the Alcock--Paczynski test, an evaluation of the ratio of observed angular size to 
radial/redshift size, is independent of it.
}

\FullConference{Frontiers of Fundamental Physics 14\\
		 15-18 July 2014\\
		 Aix Marseille University (AMU) Saint-Charles Campus, Marseille, France}

\begin{document}

\section{Does redshift mean expansion?}
\label{.redshift}

Lema\^itre\cite{Lem27} in 1927 and later Hubble\cite{Hub29} in 1929 established the 
redshift ($z$)--apparent magnitude relation of the galaxies, which gave
an observational hint that the Universe is expanding.
Hubble was cautious in suggesting this interpretation, 
but succeeding generations of cosmologists became pretty sure that
the redshift of the galaxies following Hubble's law is a definitive proof of  expansion.
There were alternative explanations for the redshifts: for instance, ``tired light'' 
scenarios\cite{Nar89}\cite{Lop03}(Sect. 2.1), in which it is assumed the photon loses energy due to 
some unknown process of photon--matter or photon--photon interaction 
when it travels some distance. This idea had two main potential problems\cite{Nar89}: the usual 
scatterings would produce blurring in the galaxies
and a frequency-dependent redshift, neither of which is observed, but  are
solved with exotic non-standard models of scattering\cite{Lop03}(Sect. 2.1).
In any case, owing mainly to the absence of a good theory based on standard physics 
explaining the possible phenomenological
fact of these alternative proposals, and given that general relativity provided an
explanation for the cosmological expansion while alternative proposals were not supported
by any well-known orthodox theory, the expansion hypothesis was preferred and  alternative
approaches were doomed to be forgotten.

\section{Observational tests for the expansion of the Universe}
\label{.testexp}

Apart from the actual redshift of the galaxies, there are different tests to verify whether the
Universe is expanding or static:

\begin{enumerate}

\item Microwave Background temperature as a function of  redshift.

The Cosmic Microwave Background Radiation (CMBR) temperature 
can be detected indirectly at high redshift if suitable absorption lines can be found in high 
redshift objects. Hot Big Bang cosmology predicts that the temperature of the CMBR required to excite 
these lines is higher than at $z=0$ by a factor $(1+z)$.

The CMBR temperature measured from the 
rotational excitation of some molecules as a function of redshift\cite{Mol02,Not11} was quite successful in proving the expansion: the results of 
Noterdaeme et al.\cite{Not11} with the exact expected dependence of
$T=T_0(1+z)$ are impressive. Nonetheless, there are other results that disagree 
with this dependence\cite{Kre12,Sat13}. The discrepancy might be due to
a dependence on collisional excitation\cite{Mol02} 
or bias due to unresolved structure\cite{Sat13}.

\item Time dilation test.

Clocks observed by us at high redshifts will appear to keep time at
a rate $(1+z)$ times slower when there is expansion. 
By using sources of known constant intrinsic periodicity,
% the light emission, 
we would expect  their light curves to be stretched in the
time axis by a factor  $(1+z)$.

Time dilation tests in Type Ia supernovae (SNIa) look like one of the most successful tests in 
favour of the expansion of the universe\cite{Gol01,Blo08}, but there are still some problems in their 
interpretation. The fact that SNIa light curves are narrower when redder\cite{Nob08} is an inconvenience 
for a clean test free from selection effects. Other selection effects and the possible compatibility of 
the results with a wider range of cosmological models, including static ones, have also been pointed 
out\cite{Lea06,Cra11}\cite{Lop03}(Secc. 2)\cite{LaV12}(Secc. 7.8). Moreover, neither gamma-ray bursts 
(GRBs)\cite{Cra09} nor Quasi Stellar Objects (QSOs)\cite{Haw10} present time dilation, which is puzzling.

\item Hubble diagram.

It has been known for  many decades that an apparent magnitude (taking into account K-corrections) vs.\ distance diagram for elliptical galaxies in clusters fits better a static rather than an expanding Universe\cite{LaV86}.
This disagreement could, however, be solved by an increase of luminosity at higher redshift due to the evolution of galaxies.

For SNIa\cite{Kow08} or GRBs\cite{Wei10}, 
for which it is supposed there is no evolution, the standard model
works, provided that an ad hoc dark energy constant is included. Nonetheless, a
static Universe may also fit those data\cite{Lop10,Mar13}. 

\item The Tolman surface brightness test.

Hubble and Tolman\cite{Hub35} proposed the so-called Tolman test based on the
measurement of the surface brightness. A galaxy at redshift $z$ varies
in surface brightness proportionally to $(1+z)^{-n}$ with $n=4$ for expansion
and $n=1$ for the static case.

Lubin \& Sandage\cite{Lub01} claimed in 2001 to have definitive 
proof of the expansion of the Universe using the Tolman test up to $z=0.9$. 
However, their claim, rather than being a Tolman test, was that the evolution
of galaxies can explain the difference between the results of the Tolman
test and their preferred model, which includes expansion. 
Lerner\cite{Ler06} observed that Lubin \& Sandage used a very involved
evolutionary k-correction scheme, with many adjustable assumptions and
parameters to correct observed high-$z$ surface brightness.
Crawford\cite{Cra11} also pointed out that Lubin \& Sandage performed a wrong analysis to exclude the static solution, mixing Big Bang and tired-light models.

Furthermore, other more recent Tolman tests\cite{Ler06,Cra11,Ler14}, some of them up to redshifts of $\sim$ 5 and with different wavelength filters so that no K-corrections are necessary, favour a static Universe without the need for galaxy evolution.

\item Angular size vs.\ redshift test. 

The angular size ($\theta $) of a galaxy with a given linear size is very different if we assume the 
standard model with expansion or a static Universe. Tests were made by 
several authors\cite{LaV86,Kap87,Lop10},
and all of them, either in the radio, near infrared or visible, show, over a range of up to redshift 3,
a dependence $\theta \sim z^{-1}$, a static Euclidean effect over all scales. 
This result cannot be reconciled with the standard cosmological model
unless we assume a strong evolution of galactic radii which coincidentaly compensates
the difference: galaxies with the same luminosity should be
six times smaller at $z=3.2$ than at $z=0$ \cite{Lop10}.
Neither the hypothesis that galaxies which formed earlier have much
higher densities nor their luminosity evolution, merger ratio,
or massive outflows due to a quasar feedback mechanism
are enough to justify such a strong size evolution\cite{Lop10}; also, 
the velocity dispersion would be much higher than observed\cite{Lop10}.
A static Universe is fitted without any ad hoc element.
However, we must be cautious with this interpretation, because 
of the uncertainty of the galaxy size evolution.

\item UV surface brightness test.

Lerner\cite{Ler06} proposed a test of the evolution hypothesis 
that is also useful in the present case. There is a limit on the ultraviolet surface 
brightness (UV SB) of a galaxy because, when the surface density of hot bright stars and
thus supernovae increases, large amounts of dust are produced to absorb 
all the UV except that from a thin layer.
Further increase in the surface density of hot bright stars beyond a given
point just produces more dust and a thinner surface layer, not an
increase in UV SB. Based on this principle, there
should be a maximum UV(at-rest) SB independent
of redshift. This was analysed
at high redshift\cite{Ler06,Lop10} and the result is that
the intrinsic UV SB would be prohibitively 
 lower(=much brighter) than 18.5 mag$_{AB}$/arcsec$^2$ 
with the evolution required for the standard model to be compatible with
the Tolman or angular size tests.
For a static model, however, it would be
within the normally expected range. Lerner\cite{Ler06} also argues why  
alternative explanations (lower production of dust at high redshift, winds 
or other scenarios) are inconsistent.
Nonetheless, Lerner's hypothesis of a maximum UV SB 
might be incorrect, so this should be further explored before
reaching definitive conclusions about this test.

\item Alcock--Paczy\'nski test.

Given a distribution of objects with spherical symmetry, with a radius along the line of sight
$s_\parallel=\Delta z\frac{d\,d_{\rm com}(z)}{dz}$
and a radius perpendicular to the line of sight
$s_\perp =\Delta \theta (1+z)^m d_{\rm ang}(z)$ ($m=1$ with expansion, $m=0$ for static),
the ratio: $y\equiv \frac{\Delta z}{z\Delta \theta }\frac{s_\perp}{s_\parallel }$
depends on the cosmological
comoving distance ($d_{\rm com}(z)$) and the angular distance ($d_{\rm ang}(z)$)
and is independent of the evolution of galaxies, 
but it also depends on the redshift distortions produced by the peculiar velocities of  gravitational infall\cite{Lop14}. 

L\'opez-Corredoira\cite{Lop14} measured $y(z)$ by means of the
analysis of the anisotropic correlation function of sour\-ces in several surveys, using a technique to disentangle the dynamic and geometric distortions, and also took other values available from the literature. From six different cosmological models (concordance $\Lambda $CDM, Einstein-de Sitter, open--Friedman Cosmology without dark energy, flat quasi-steady state cosmology, a static universe with a linear Hubble law, and a static universe with tired--light redshift), only two of them fit the data of the Alcock \& Paczy\'nski's test: concordance $\Lambda $CDM and static universe with tired-light redshift; whereas the rest were excluded at a $>95$\% confidence level. Analyses with further data using Baryonic Acoustic 
Oscillations (BAO) improve the test and give us a more accurate constraint, but do not
exclude neither the static case nor expansion yet\cite{Lop15}.

\end{enumerate}

\section{Conclusions}

\begin{table*}
\caption{Cosmological tests.}
\begin{center}
\begin{tabular}{ccc}
\label{Tab:tests}
Test & Expansion & Static \\ \hline \hline
$T_{\rm CMBR}(z)$ & Good fit & $\begin{array}{l}
{\rm Excess\ temperature\ at\ high\ z} \\
{\rm due\ to\ collisional\ excitation\ or} \\
{\rm due\ to\ unresolved\ structure.}
\end{array}$ \\ \hline
Time dilation & $\begin{array}{l}
{\rm Good\ fit\ for\ SNIa.} \\
{\rm Unexplained\ absence\ of\ time} \\
{\rm dilation\ for\ QSOs\ and\ GRBs.} 
\end{array}$ & $\begin{array}{l}
{\rm Selection\ effects,\ or\ ad\ hoc} \\
{\rm modification\ of\ the\ theory\ or} \\
{\rm the\ zero\ point\ calibration,\ or} \\
{\rm evolution\ of\ SNIa\ periods.}
\end{array}$ \\ \hline
Hubble diagram & $\begin{array}{l}
{\rm Requires\ introduction\ of\ dark} \\
{\rm energy\ and/or\ evolution.}
\end{array}$ & $\begin{array}{l}
{\rm Good\ fit\ for\ galaxies.\ Good} \\
{\rm fit\ for\ SNIa\ with\ some\ models.}
\end{array}$ \\ \hline
Tolman (SB) & Requires strong SB evolution. &
Good fit \\ \hline
Angular size & $\begin{array}{l}
{\rm Requires\ too\ strong}\\
{\rm evolution\ of\ angular\ sizes.} 
\end{array}$ & Good fit \\ \hline
UV SB limit & Too high UV SB at
high z & Within the constraints \\ \hline
Alcock-Paczy\'nski & Good fit & Good fit for tired light \\ \hline \hline
\end{tabular}
\end{center}
\end{table*}

Table \ref{Tab:tests} summarizes the analysis of this paper. Apparently, there is no winner
yet. The first two tests favour expansion, whereas the following four tests get 
a less ad hoc fit with the static solution, although this is insufficient to reject expansion. Most of the cosmological tests are entangled with the evolution of galaxies and/or other effects. Tolman or angular size tests need to assume a very strong evolution of galaxy sizes to fit the data with the standard cosmology, whereas the Alcock--Paczynski test is independent of the evolution of galaxies.

\end{document}